%% file: articleShell.tex
\documentclass[twocolumn]{classes/smallSat} 
\journal{basic}
\articletype{Proceedings}

\def\doctitle{Open-Source CubeSat Solar Panels: Design, Assembly, Testing, and On-Orbit Demonstration}
\def\authorOne{Nicholas J. Sorensen\orcidlink{0000-0002-4666-5791}}
\def\authorTwo{Erik F. Halliwell\orcidlink{0009-0002-7510-6670}}
\def\affilOne{Dept. of Physics and Astronomy, University of Calgary}
\def\affilTwo{Dept. of Electrical and Computer Engineering, University of Alberta}
\def\addressOnlyOne{834 Campus Pl NW, Calgary AB T2N 1N4, Canada}
\def\addressOnlyTwo{9211 116 St NW, Edmonton AB T6G 1H9, Canada}
\def\phoneOne{+1 (780) 996-5563}
\def\phoneTwo{+1 (780) 667-1168}
\def\emailOne{nicholas.sorensen@ucalgary.ca}
\def\emailTwo{erik.halliwell@ualberta.ca}

\def\paperNumber{SSC24-WP2-33} 

\def\abstractText{
\noindent Cube satellites, or CubeSats, are small satellites commonly used to perform Earth imaging and on-orbit scientific experiments. 
CubeSats are often powered using expensive, inflexible commercial-off-the-shelf solar panels, largely due to a lack of flight-qualified open-source alternatives.
Here, we describe the design of customizable, deployable solar panels, offering an open-source, cost-effective alternative.
Towards a fully open-source CubeSat, our designs have mission-tailored power generation capabilities and simple electrical and mechanical integration.
The solar panel designs were demonstrated on-orbit on three satellites in the Northern SPIRIT constellation and will be on AlbertaSat's Ex-Alta~3 satellite, which will launch in 2025. 
The design files, assembly procedures, and best practices will be open-source-published online. 
This work lowers the barrier of entry into space, making satellite design easier and less expensive - students helping students design better satellites.
} 

\usepackage{orcidlink,breqn,enumitem,caption,subcaption,ulem,titlesec,lipsum,dirtytalk,mathtools,gensymb,float,setspace,siunitx}
\usepackage{multirow}
\usepackage[nameinlink]{cleveref}
\usepackage{soul}

\begin{document}

\twocolumn[
\input{content/articleFrontMatter} 
]
\input{content/articleMain}

\bibliography{bibliographies/articleShell}

\end{document}

%% file: content/articleFrontMatter.tex
\title{\doctitle}

\author{
    \customauthor{\authorOne}{\affilOne}{\addressOnlyOne}{\phoneOne}{\emailOne} \\
    \vskip 1em
    \customauthor{\authorTwo}{\affilTwo}{\addressOnlyTwo}{\phoneTwo}{\emailTwo}
}
{\raggedleft\Large\textbf{\paperNumber}\par}

\maketitle
\begin{abstract}
\vspace{-10pt}
\abstractText
\vspace{5pt}
\end{abstract}

\date{}

%% file: content/articleMain.tex
\section*{Introduction}
\vspace{-1em}
Cube satellites, also referred to as CubeSats, are small modular satellites used to perform Earth observation~\cite{Ghuffar2018, Jervis2021}, scientific research~\cite{Mann2020}, and satellite communication~\cite{Zheng2020}, among other applications.
They are categorized by the number of units
they comprise, where one unit (1U) is
a ($10 \times 10 \times 10$) cm$^3$ module.
Due to their relative inexpensiveness and simplicity, the number of CubeSats in orbit is steadily increasing \cite{Villela2019}.
Most CubeSats are powered using solar panels, often purchased as commercial-off-the-shelf (COTS) products.
This can partly be attributed to a lack of flight-qualified open-source alternatives.
To address this need, we present the design for inexpensive, customizable, deployable solar panels toward a fully open-source satellite.\\

\noindent The use of COTS solar panels on CubeSats has several disadvantages. 
Firstly, COTS solar panels are electrically and mechanically challenging to integrate with payloads, instrumentation, and other satellite bus components.
Further, standard COTS body-mounted solar panels for a 3U CubeSat can cost tens of thousands of USD - without customization options or deployables.
While a significant portion of this is material cost, the cost of labor, development, and testing contribute significantly.\\
\begin{figure}[t]
    \centering
    \includegraphics[width = 0.49\textwidth]{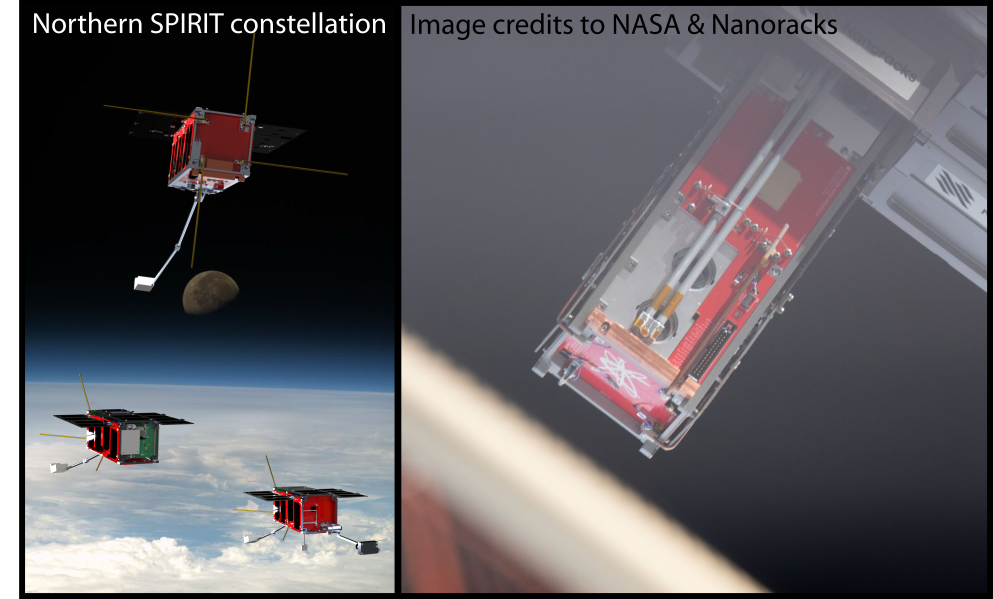}
    \caption{The Northern SPIRIT constellation}
    \label{fig:rendersImages}
\end{figure}

\noindent Using in-house, customizable panels can solve these problems, particularly for student group projects.
Given the necessary design files and procedures, student volunteers can manufacture and fly in-house panels - reducing their associated costs.
Further, in-house manufacturing trains students and creates highly qualified personnel.
These are often some of the primary objectives of student CubeSat missions. 
Customizable solar panels also have design advantages.
Namely, unsegmented panels add structural integrity, reduce the complexity of deployment mechanisms, and can help maximize the utility of other bus systems, such as the electrical power system (EPS) or communications system.
Moreover, custom panels permit easier mechanical integration and a mission-tailored power generation solution.
Therefore, we produced a customizable solar panel design for Ex-Alta~2 - the University of Alberta’s second satellite. \\
\vspace{-1.8em}
\subsection*{The Northern SPIRIT CubeSat Constellation}
In 2017, AlbertaSat, an undergraduate student-led project from the University of Alberta, launched the first Experimental Albertan \#1 (Ex-Alta~1) satellite.
It performed magnetometry experiments in low-earth-orbit (LEO) \cite{Mann2020} as part of the QB50 CubeSat network \cite{Masutti_2012}.
Following its success, AlbertaSat developed a constellation of satellites (renders shown in \Cref{fig:rendersImages}, left) as part of the Northern Space Program for Innovative Research and Integrated Training (SPIRIT) in collaboration with two other student groups at Yukon University in Whitehorse, YK, and the Aurora Research Insitute in Inuvik, NWT.
The constellation was funded through the Canadian Space Agency's Canadian CubeSat Project (CCP) and consisted of three satellites: Ex-Alta~2 (\Cref{fig:rendersImages}, right), a 3-unit (3U) CubeSat, and YukonSat and AuroraSat, two 2U CubeSats. 
One mission objective of Ex-Alta~2 was to perform wildfire imaging and science using an in-house imaging system.
Another objective of the constellation was to produce solar panels for the three satellites.
Open-source solutions were largely unavailable, so the student team invested significant time and resources into developing custom, in-house solar panels \cite{Halliwell_2018}.
The solar panels flew on the Ex-Alta~2, AuroraSat, and YukonSat CubeSats, deployed from the ISS on April 24, 2023, as part of NanoRacks NRCSD25.
We believe the solar panels functioned nominally in orbit, and we discuss their design in this paper.\\
\vspace{-1.8em}
\subsection*{Paper Outline}
Here, we present the customizable design of a solar panel array towards a fully open-source CubeSat.
We describe the solar panel design and assembly procedures, test results, and plans for the future.
Before launching the three CubeSats, the panels underwent thermal vacuum (TVAC) and vibration qualification testing and performance characterization using state-of-the-art solar simulation equipment.
We present the results of these tests and discuss analyses of potential deployment failure mechanisms.
Further, we describe the design of solar panel-integrated magnetorquers, which can replace torque rods used to control the orientation of the satellites~\cite{Sorensen2021}.\\

\noindent Our solar panel designs will also be used on AlbertaSat’s Ex-Alta~3 CubeSat, scheduled for launch in 2025. 
Additionally, we will publish the design files, assembly procedures, and best practices online under the Apache 2.0 license, furthering the goal of a fully open-source CubeSat.
This work lowers the barrier of entry to designing spacecraft, simplifying CubeSat design - students helping students build satellites.
\begin{figure*}[t]
    \centering
    \includegraphics[width = \textwidth]{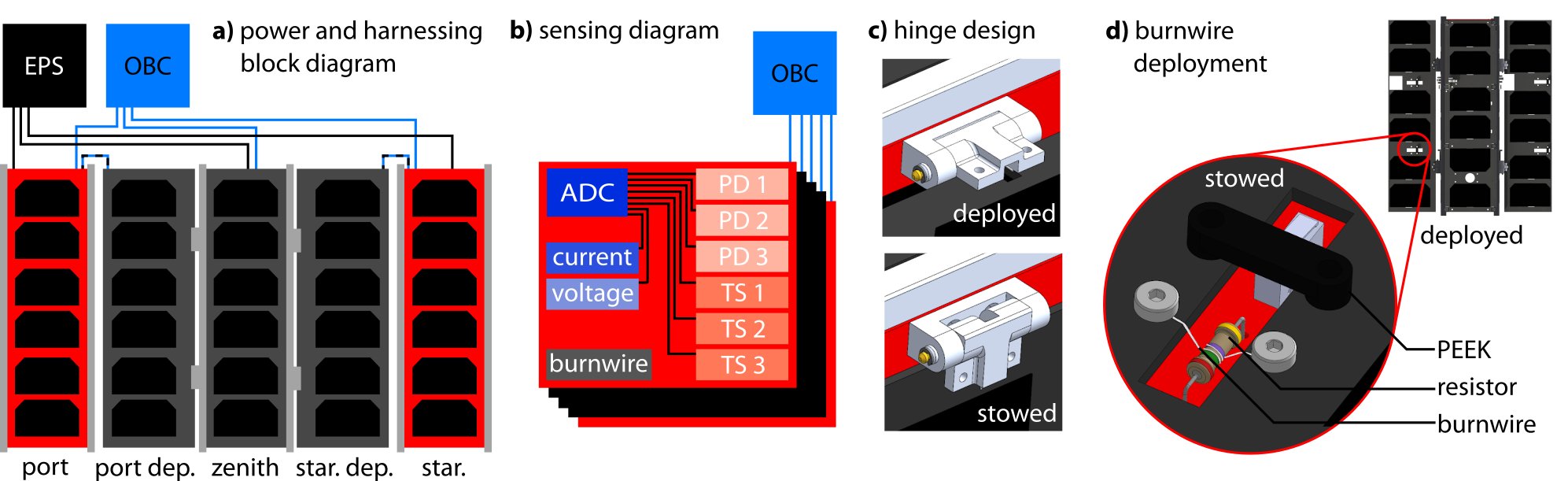}
    \caption{Power and sensing block diagrams of the solar panels, and a burnwire deployment mechanism sketch}
    \label{fig:figBlockDiagramsDesign}
\end{figure*}
\vspace{-0.5em}
\section*{Solar panel design}
All three Northern SPIRIT satellites used solar panels designed in a wing-deployment configuration, which consistently generated enough power to meet the substantial power demands of the satellite payloads.
Ex-Alta~2 utilized five individual solar panels, each housing six of Spectrolab's GaAs XTJ-Prime solar cells (see \Cref{fig:figBlockDiagramsDesign}a).
The 2U satellite panels used two or four solar cells selected for their favorable efficiency, small size, and integrated bypass diodes.
The satellites were earth-facing, prompting us to use three body-mounted panels: one on zenith, one on port, and one on starboard. 
Additionally, there were two deployable panels; when stowed during launch, one panel was bound to port and the other to starboard using burnwires. 
When deployed, each panel sat parallel to zenith (see \Cref{fig:rendersImages}).  \\

\input{tables/electricalCharTable}
\noindent On Ex-Alta~2, the cells were arranged in a six-series one-parallel (6S1P) configuration, nominally producing \SI{14.4}{V} at \SI{0.48}{A} at the maximum power point (MPP). 
We summarize some of the electrical and mechanical solar panel characteristics of Ex-Alta~2 in \Cref{tab:char}.
The 2U solar panel characteristics follow similar trends, most being smaller by a factor of two-thirds.\\

\noindent Each panel integrated several different sensors, including a current sensor, a voltage sensor, three photodiode sensors, and three temperature sensors (see \Cref{fig:figBlockDiagramsDesign}b). 
Each sensor interfaced with an ADC, which communicated with the on-board computer (OBC) over I2C. 
Additionally, the panels housed sun sensors used by the attitude determination and control system (ADCS). \\

\subsection*{Deployment Mechanisms}
The deployable solar panels had several important deployment mechanisms to deploy different components, including the panels themselves.
The deployable panels were attached to the body-mounted panels using aluminum hinges with springs, designed by a mechanical engineering undergraduate capstone project (see \Cref{fig:figBlockDiagramsDesign}c). 
These hinges were machined in-house and anodized alongside the rest of the mechanical chassis. 
The panels were actuated using burnwires similar to those described by Damkjar et al.~\cite{Damkjar2019}, and \Cref{fig:figBlockDiagramsDesign}d shows a sketch of the deployment mechanism; the ultra-high-frequency (UHF) antennae and the magnetometer boom were deployed using similar mechanisms. 
When initiating deployment, the OBC actuated \SI{5}{V} to the \SI{10}{\ohm} flame-resistant metal film (part number CPF110R000FKEE6) burnwire resistors. 
The resistors heated up to approximately \SI{200}{\degree C}, melting the twice-wrapped Dyneema thread (Berkley Solutions Braid, BSBFS10-22) used to secure the deployable panels.
To sense the deployment state of the components, we used single-pull, double-throw (SPDT) switches (PANA-AV4424).
Further, the deployment switches and burnwire mechanisms were designed into the body-mounted panels to work around mechanical constraints using appropriate cutouts to make electronic harnessing easier.
Several 3D-printed PEEK parts were used in the deployment mechanisms.\\
\vspace{-2em}
\section*{Solar panel assembly methods}
\begin{figure*}
    \centering
    \includegraphics[width = \textwidth]{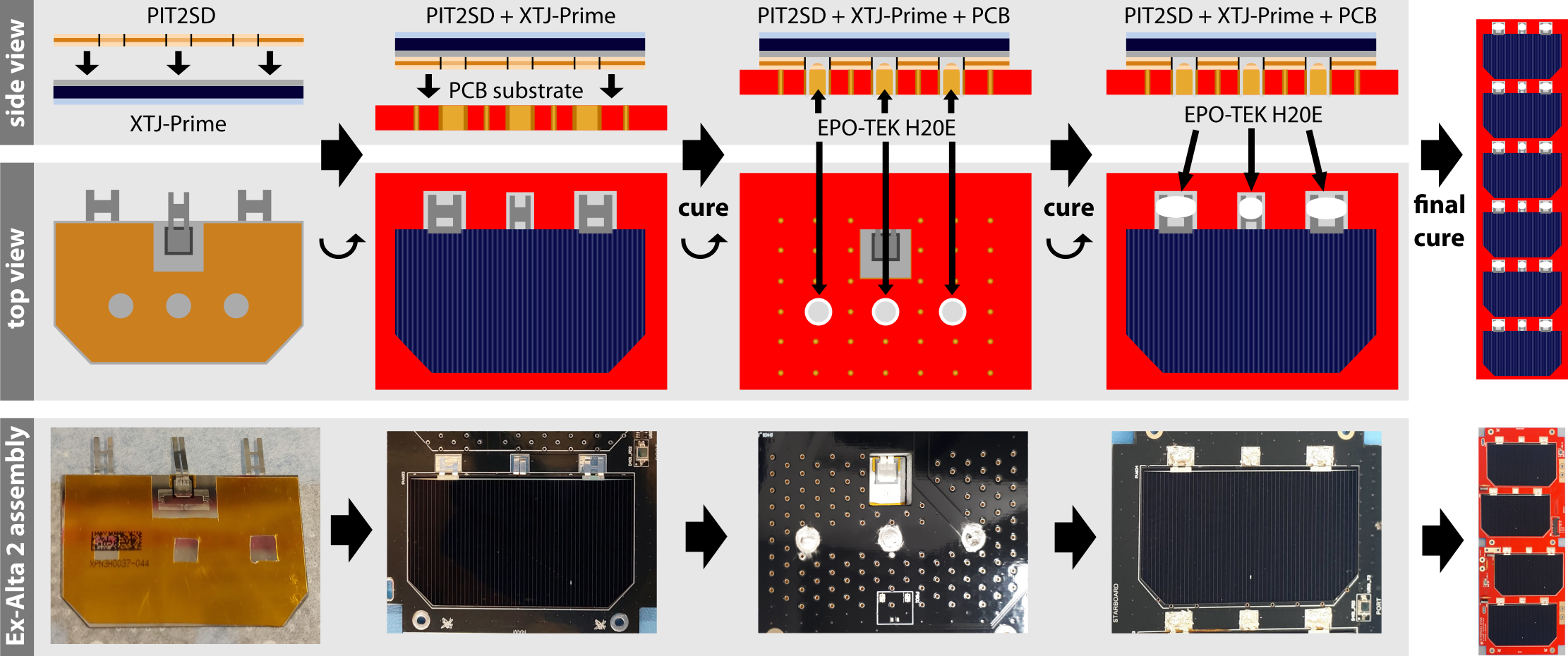}
    \caption{Process used to assemble Northern SPIRIT solar array}
    \label{fig:panelAssembly}
\end{figure*}
This section describes how we adhere the solar cells to the printed circuit board (PCB) substrates.
Our method was built upon several previous works~\cite{Vanhille_2000, Karuza_2009, Keller_2012, Sandberg2016, Dahir2017, Lisy2023}; to attach the cells to the substrate, Keller et al., \cite{Keller_2012} used a conductive epoxy, whereas Vanhille \cite{Vanhille_2000} and Karuza et al. \cite{Karuza_2009} used two layers of Kapton tape. 
Each method had several drawbacks.
Keller's approach is used widely by student CubeSat projects to assemble solar panels; however, it can cause the formation of air pockets or voids between the solar cells and the substrates. 
When exposed to vacuum, these air pockets can crack the solar cells, reducing their potential to generate power. 
Karuza's method showed improvement in reducing voids and was flight-qualified on AeroCube-3. 
Their technique uses two layers of Kapton tape - one attached directly to the substrate and the other - double-sided - attached the cell to the substrate.
Two layers of Kapton tape still permit the formation of voids within the layers, which can damage the solar cells in vacuum. \\

\noindent To reduce the risk of creating air pockets - also called voids - two strategies were adopted. 
Firstly, we use a single piece of double-sided Kapton tape, adhered first to the solar cells.
This method is more repeatable than using conductive epoxy, and it reduces the risk of forming voids.
Secondly, we add via stitching to the solar panel PCB substrates.
This via stitching, if sufficiently dense ($\gtrapprox\SI{5}{cm^2}$), nearly eliminates the possibility of air pockets forming between the tape and the substrate. 
Procedural details on how we assemble the panels using these methods are given in the next section.\\
\vspace{-1.8em}
\subsection*{Initial Panel Assembly and Verification} 
An early version of these panels was created for the fourth Canadian Satellite Design Challenge (CSDC) student competition. 
The cost of space-grade triple junction gallium arsenide solar cells was prohibitive, so we tested an early iteration of the cell assembly procedure using silicon solar cells. 
The silicon cells, which initially measured $\SI{80}{mm}\times\SI{80}{mm}$, were mechanically scoured with a scalpel and snapped along the growth direction crystal axis to be the same dimensions as the Ex-Alta~2 XTJ prime solar cells - $\SI{69}{mm}\times\SI{40}{mm}$.  These cells were then attached to the PCBs using similar methods to those described in the following section.\\

\noindent The panels were inspected for voids using a thermal camera both before and after vibration testing, which was conducted at the Canadian Space Agency (CSA) David Florida Laboratory complex. 
The success of the tests permitted us to design and build the solar panels for the Northern SPIRIT satellites.
A summary of this preliminary work was presented in 2018 at the 10th European CubeSat Symposium~\cite{Halliwell_2018}.\\

\noindent The solar cells are attached to the PCB using a double-sided polyimide (Kapton) tape, which uses a silicone adhesive, and the electrodes are connected using a silver epoxy. 
Specifically, we used the CAPLINQ PIT2SD, \SI{5}{mil}-thick tape, and EPO-TEK H20E silver epoxy. 
The utilized tape had higher total mass loss and collected volatile condensable material than specified by the mission requirements; however, the small amount of exposed surface area of the tape let us deviate from the mission requirements. 
For projects with stricter requirements, we recommend using NuSil CV4-1161-5 double-sided polyimide tape.
The panels are assembled after all other electronic components have been added and fully tested; this includes all sensing and burnwire electronics. \\

\begin{figure*}
    \centering
    \includegraphics[width=\textwidth]{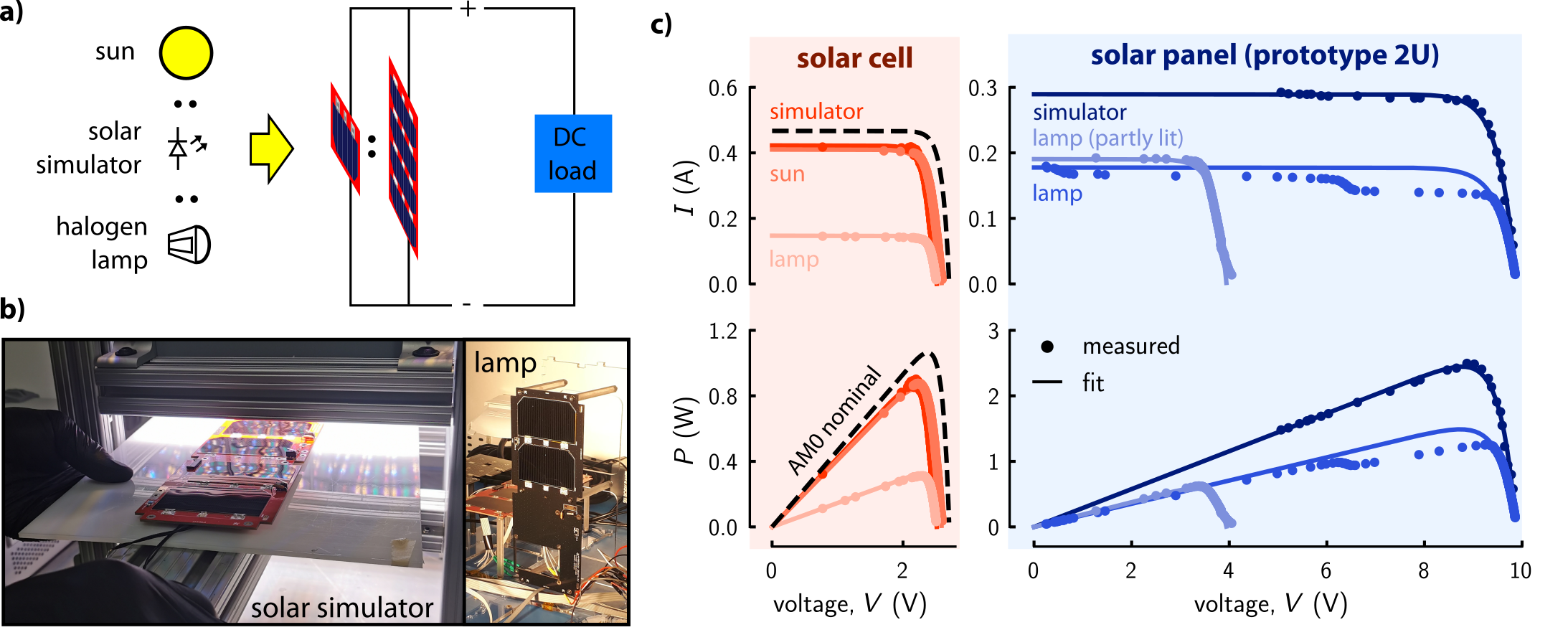}
    \caption{Characterization of solar cell and solar panel power generation under different illumination schemes}
    \label{fig:solarCellChar}
\end{figure*}
\vspace{-2em}
\subsection*{Final Assembly Methods} 
Now, we describe the procedure used to attach the solar cells, shown in \Cref{fig:panelAssembly}. 
First, the tape is cut to the same dimensions as the solar cells using a water-cut, aluminum template.
Holes are cut in the tape, mirroring the three large holes and cutouts on the PCB substrate, used to electrically connect the positive terminal of the solar cells and give room to the bypass diodes.
The tape is applied to the solar cell and smoothed to remove any air bubbles; this is done carefully on a flat, clean surface - the cells are extremely fragile and can break easily.
Then, the tape/solar cell assembly is carefully applied to the substrate.
To avoid trapped air we place venting holes in the PCB substrate.
The silver epoxy is mixed as per the manufacturer's recommendations and is then off-gassed in a \SI{-27}{mmHg} vacuum chamber for ten minutes. 
A small amount of the epoxy ($\sim\SI{0.25}{mL}$) is applied to the three large holes on the underside of the PCB; if too much epoxy is used, it will degas and expand as it cures, which can cause significant stress on the solar cells.
Then, the epoxy is cured at \SI{120}{\degree C} for two hours in a partial vacuum (\SI{-10}{mmHg}), pausing halfway to allow for electrical short testing.
Once complete, the panels are allowed to cool at room temperature.
Epoxy is then added to the negative terminals on the top of the panel, again using only a small amount, before curing at \SI{100}{\degree C} for two more hours in the same partial vacuum (\SI{-10}{mmHg}).\\

\noindent Once the assembly is complete, solar array testing may commence.
All component-level testing is done before adhering the solar cells to the substrates, which reduces the likelihood of cell damage.
This assembly method was verified during system and vibration tests of the Northern SPIRIT constellation.\\
\vspace{-1em}
\section*{Panel verification and testing}
\begin{figure}[htb]
    \centering
    \includegraphics[width = 0.48\textwidth]{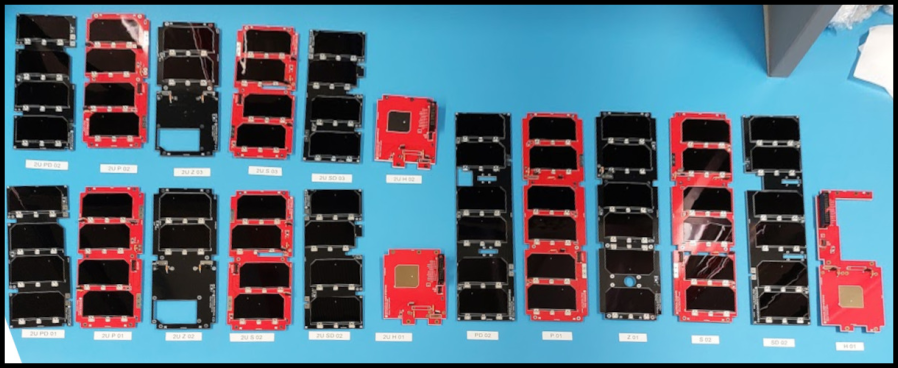}
    \caption{Assembled flight-model solar panels for the Northern SPIRIT constellation}
    \label{fig:protoflight}
\end{figure}

There were several rounds of prototyping and testing (summarized in \Cref{fig:solarCellChar}), however, in this section, we focus on testing done to the flight model panels, shown in \Cref{fig:protoflight}.
Before the cells were attached, all other components were soldered and tested.
Relevant tests included continuity checks, power consumption, characterization of the current and voltage sensors, temperature and photodiode sensor testing, and I2C bus tests. \\

\noindent Then, all solar cells were tested individually. 
Each cell was visually inspected for abnormalities, weighed, and dimensioned.
Some of the cells also underwent electrical performance measurements, during which we measured the voltage-dependent power generation curves of the cells.
A Rigol DL3021 programmable DC load was used to measure the current and voltage produced by a solar cell for different load levels under various illumination levels. 
To test the solar cells, we used several different light sources, including a \SI{15}{W} halogen lamp, the sun on a sunny day (in Edmonton, equivalent to air mass 1.2 [AM1.2]), and a G2V LED Sunbrick Solar Simulator (AM1.2) (see \Cref{fig:solarCellChar}a). Next, we characterized the voltage dependence of current and power generation for a fully lit panel and a partially lit panel; we achieved partial illumination by covering half of the solar panel, and this was done to verify bypass diode functionality. 
In \Cref{fig:solarCellChar}b we show two images from testing using the G2V solar simulator (left) and the halogen lamp (right). \\
In \Cref{fig:solarCellChar}c we plot the current and power characterization curves for the solar cell and the prototype 2U panel, and we compare the solar cell results to that expected at AM0, or in low earth orbit. 
The power curves of the prototype panels met performance expectations at AM1.2, but the maximal power generation at AM0 was 30\% lower than expected. 
This reduced performance was caused by incidental contact with one of the prototype panel's cells during assembly.
This damage results in the stepping seen in \Cref{fig:solarCellChar}c, caused by activation of the bypass diodes. 
Stepping can also signify uneven illumination of the panels, which reduces the total power generation.
Some cells underwent reverse bias testing to identify any points of damage.
Subsequent models, including the panels used for flight, were proven through testing to be damage-free and behaved nominally. \\

\begin{figure}[htb]
    \centering
    \includegraphics[width = 0.48\textwidth]{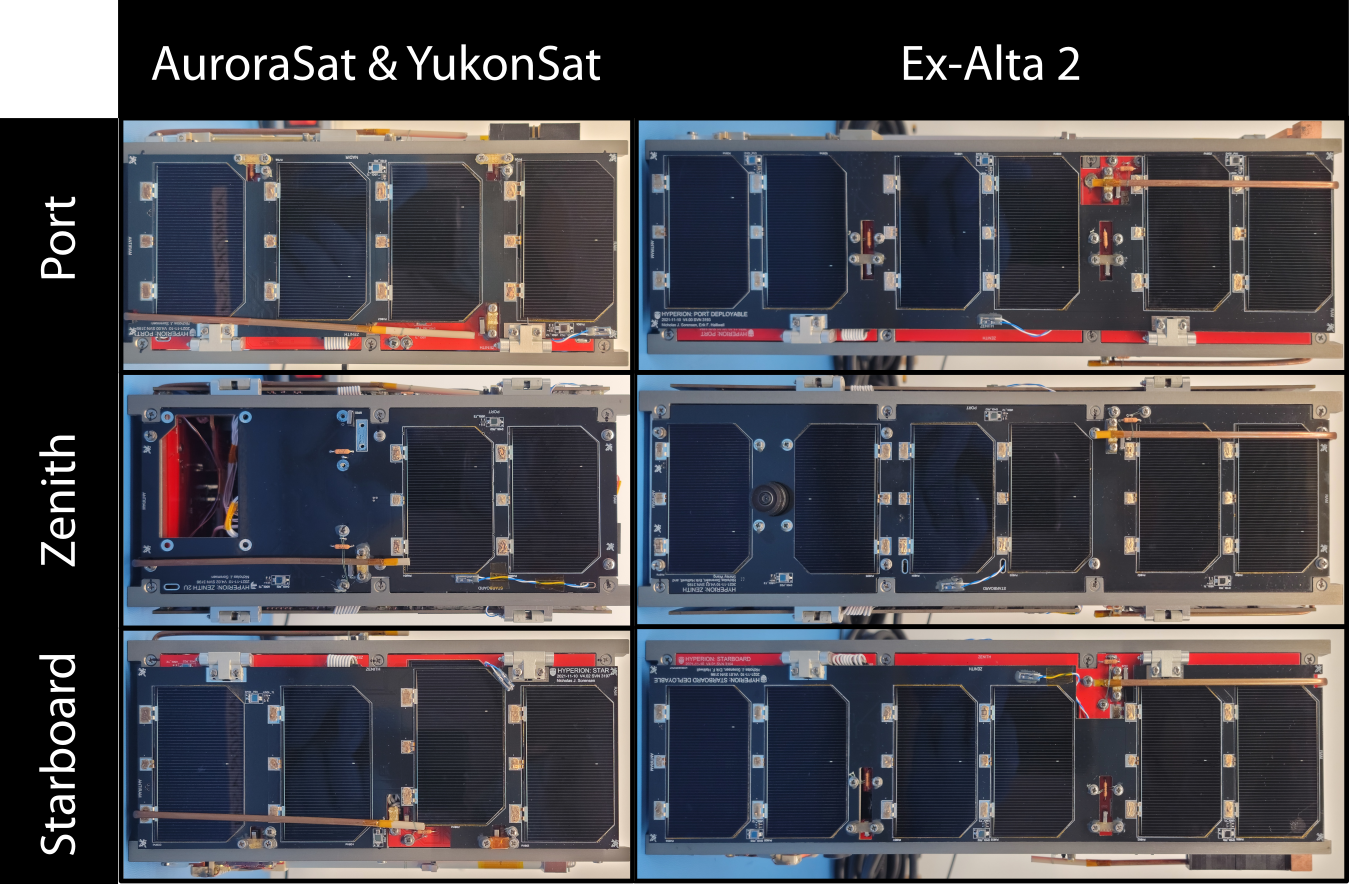}
    \caption{Assembled flight-model CubeSats}
    \label{fig:FlightModelSats}
\end{figure}

\begin{figure}[htb]
    \centering
    \includegraphics[width=0.49\textwidth]{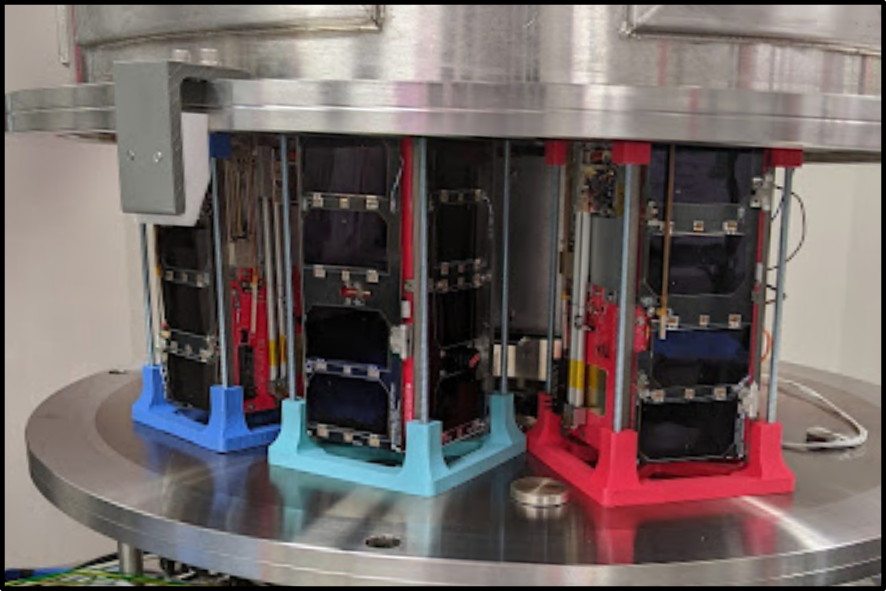}
    \caption{Thermal-vacuum testing of the three CubeSat assemblies}
    \label{fig:TVAC}
\end{figure}

\noindent Beyond testing the panels individually, the solar panels also underwent TVAC, subsystem, and system-wide tests. 
The electrical and sensing characteristics of the assembled solar panels were tested again after connecting them to the EPS and OBC.
The satellite assemblies were held at high vacuum ($<10^{-6}$ Torr) for six hours in a TVAC chamber, followed by full functional tests of the entire satellite assembly (\Cref{fig:FlightModelSats}). 
The TVAC chamber (\Cref{fig:TVAC}) was designed in part by student group members in AlbertaSat and by a mechanical engineering undergraduate capstone project. 

\vspace{-2em}
\section*{Impact of Deployment Failure}
In this section, we discuss the potential impacts of solar panel deployment failure on satellite functionality.
As described earlier, the solar panels housed several deployment mechanisms to deploy different satellite components, including payloads, UHF antennae, and the panels themselves.
Despite attempts to mitigate risk, such mechanisms could fail to deploy.
Therefore, we had to assess the consequences of such failures.
To do so, we considered two deployment failure configurations and assessed their impact on system functionality, primarily considering the system power constraints.
The presented analysis methods can be performed for any solar panel or deployable configuration.\\

\noindent We considered two failure configurations: the nominal configuration constituted the successful deployment of all components, and the failure configuration constituted the deployment failure of both deployable panels.
By failing to deploy, the body-mounted panels would not generate any power, affecting the operational power budget of the satellite.\\

\noindent To understand the potential impact of deployment failure on satellite operation, it was critical to know the power requirements of the satellite in different operational modes.
As satellites orbit, they perform different tasks, such as payload operation and data downlink.
An operational mode is defined by the task or series of tasks performed over time, and each mode requires different amounts of power.
We can compare the amount of power required to operate in a given mode to that generated by the satellite over a given period, which allows us to assess satellite functionality (i.e. how frequently the satellite can perform measurements or tasks).
We considered three operational modes: a critical mode, when the satellite operated at minimum capacity; a standby mode, when the satellite performed nominal computation and communication tasks, and a full-function mode, when the satellite executed payload tasks in addition to those listed in the nominal mode.
The satellite consumed different amounts of power in each mode, given by the operational characteristics of the bus components.
We used these associated powers to assess satellite functionality in different failure configuration schemes.
Ex-Alta~2 consumed \SI{3.8}{W} nominally, and \SI{6.1}{W} when fully functioning. 
When the EPS reached critical mode, the satellite consumed as little as \SI{20}{mW}. \\

\begin{figure}[htb]
    \centering
    \includegraphics[width = 0.48\textwidth]{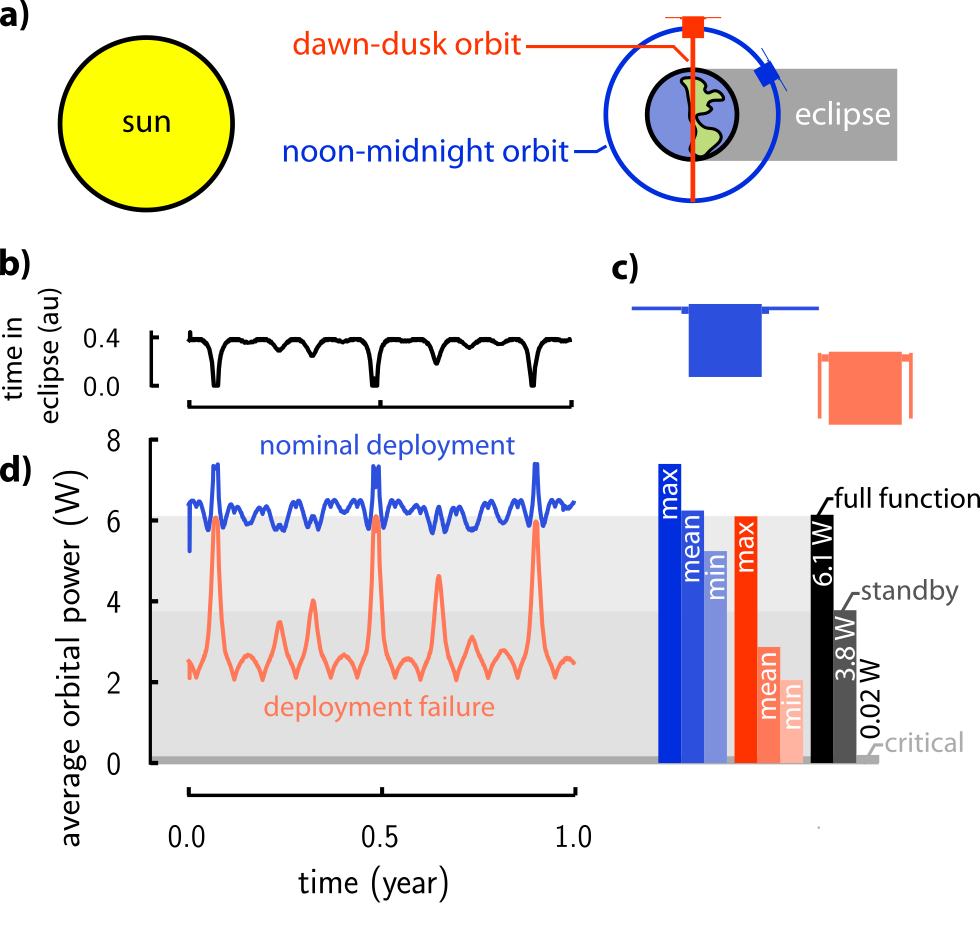}
    \caption{Simulation of satellite power generation}
    \label{fig:figFailureAnalysis}
\end{figure}

\noindent To characterize satellite power generation in each failure configuration, we performed simulations in STK.
Using different three-dimensional models for each failure configuration, STK simulated the orbit of each satellite.
Specifically, the simulation used an SGP4 propagator, which considered secular and periodic variations due to Earth’s oblateness, solar and lunar gravitational effects, and orbital decay.
We simulated the orbital progression of the satellite over one year and calculated the proportion of the orbit spent in eclipse.
This parameter was periodic, which resulted from the precession of the orbital plane of the satellite between a dusk-dawn orbit (umbra minima) and a noon-midnight orbit (umbra maxima), shown in \Cref{fig:figFailureAnalysis}a. 
The proportion of the orbit spent in eclipse is plotted in \Cref{fig:figFailureAnalysis}b.
This result simulation demonstrated that, even in a worst-case LEO, the solar panels generate power for at least 60\% of any orbit.
Next, we simulated the power generation of each panel for all three satellites in their nominal and failure deployment configurations (\Cref{fig:figFailureAnalysis}c) over one year.\\

\noindent In \Cref{fig:figFailureAnalysis}d, we plot the per orbit-averaged total power generation of Ex-Alta~2 for both configurations over one year.
We see that in the nominal configuration, the solar panel design produced a consistent supply of power (average, $\SI{9.6}{W} \pm 17$\%).
This small amount of variation in power generation, despite complex orbital dynamics, indicated good design.
In comparison, the failure configuration produced lower average power (\SI{2.9}{W}), with a much greater orbital variation.
Recall that in this configuration, the two deployable panels have failed to deploy, rendering the Port and Starboard body-mounted panels useless. \\

\noindent Comparing the generated power in each configuration to the expected power consumption (\Cref{fig:figFailureAnalysis}d, right), it is clear that the satellite achieved full functionality in a nominal deployment configuration.
In fact, the mean power generation exceeded the required power at full function.
In the deployment failure configuration, however, functionality would be reduced.
The mean power generation in that case was less than the standby mode power requirement, and the functionality or duty cycle of the satellite would need to be reduced to maintain power positivity. 
In either failure configuration, the satellite would still function, as the critical power requirement is much lower than the worst power generation case. \\

\noindent For different satellites and solar panel designs, similar analyses should be performed to determine the effectiveness of the solar panel design. 

\vspace{-1em}
\section*{Integrated Magnetorquer design}
In this section, we briefly review the design of magnetorquer coils embedded in prototype solar panels of the Northern SPIRIT satellites.\\

\begin{figure}[htb]
    \centering
    \includegraphics[width = 0.45\textwidth]{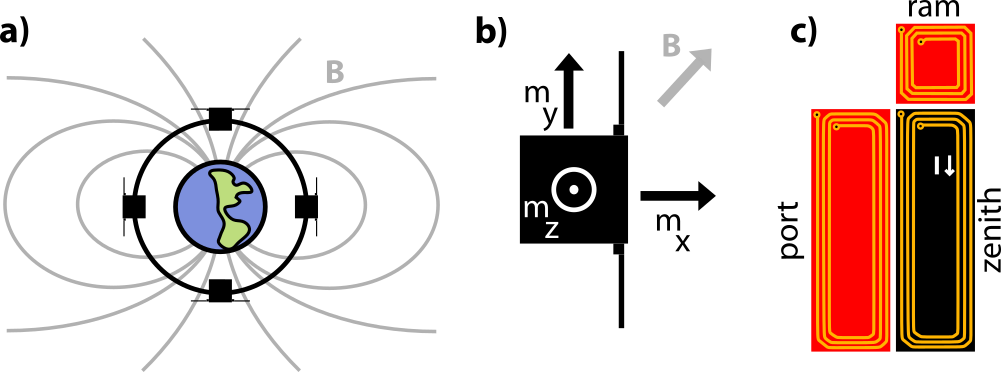}
    \caption{Solar panel-embedded magnetorquers}
    \label{fig:figMagnetorquers}
\end{figure}

\noindent Often, it is vital to precisely control the attitude or orientation of a satellite as it orbits (see \Cref{fig:figMagnetorquers}a). 
Attitude control is typically achieved using an ADCS, which can be challenging to design and integrate\cite{Chen1999}. 
Typical ADCS systems control the spacecraft's attitude using gyroscopic torque wheels and magnetorquer rods, both of which can be bulky and can consume a large portion of the volume of a CubeSat~\cite{Chen1999}.
Magnetorquer rods constitute wire loops that conduct electric currents.
These current loops produce tunable magnetic dipole moments that interact with Earth's magnetic field and rotate the satellite~\cite{Grau2017}. 
Using three magnetorquer coils placed on orthogonal planes of the satellite, it is possible to control its orientation fully (see \Cref{fig:figMagnetorquers}b). \\

\noindent To reduce the volume requirements of an ADCS, magnetorquer coils can be integrated into other satellite systems, such as solar panels.
They consume much less volume compared to standalone COTS magnetorquers, however, they tend to be less efficient ($10\sim 30\%$ as efficient, depending on the size of the panel)\cite{Sorensen2021}.
Several investigations have studied solar panel-embedded magnetorquers and produced optimized design methods\cite{Sorensen2021, Ali2012, Musiab2016, Ali2021}.
In particular, Sorensen produced a method to optimize the efficiency of magnetorquers embedded into CubeSat solar panels, and, using this methodology, magnetorquers were embedded into several prototype iterations of the North SPIRIT solar panels~\cite{Sorensen2021}.
The magnetorquers were not integrated into the flight model panels, as the satellites used separate COTS ADCS systems, deeming the embedded magnetorquers unnecessary. \\

\noindent The prototype magnetorquers were integrated into three of the body-mounted panels, each on a different plane of the satellite: ram, zenith, and port (see \Cref{fig:figMagnetorquers}c). 
Their geometries were optimized to produce the greatest magnetic moment for different voltage and power consumption levels.
Specifically, each magnetorquer was designed to consume \SI{1}{W} at \SI{3.3}{V}, producing a magnetic moment of \SI{0.22}{A m^2} for the starboard/port and zenith panels, and \SI{0.08}{A m^2} for the ram/anti-ram panel~\cite{Sorensen2021}.
Different magnetic moments were achieved by varying the magnetorquer duty cycles.
Efficiency optimizations were computed using COMSOL for designs using only a single conductive layer of the solar panel PCB.
Higher efficiencies could be achieved using thicker PCBs with more conductive layers.
Precise optimization of PCB layout, however, is not necessary to produce an effective design; efficiency generally increases with the number of conductive loops and the surface area enclosed by each conductive loop, and the power consumption can be tailored by changing the cross-sectional area and length of the wire traces.

\section*{Conclusion and future work}
We have presented the design, assembly, and testing of customizable, deployable solar panels flown on satellites in the Northern SPIRIT constellation. 
Our designs address the limitations of commercial-off-the-shelf solutions, offering an open-source, cost-effective alternative that eases electrical and mechanical integration, improves structural integrity, and allows for better-tailored power generation capabilities. 
The design and test methodology presented in this article can be adopted by other CubeSat mission teams, especially those made by student project teams. \\

\noindent We outlined the design's power generation and sensing capabilities and the requirements that motivated many design decisions.
We also illustrated how our designs could be altered to meet the requirements of future satellite missions.
Further, we detailed the assembly procedures for the solar panels, describing the methods and materials used to electrically and mechanically attach the solar cells to their PCB substrates.
Characterization methods to verify the solar panel performance were detailed. 
These characterizations included electrical performance tests of the individual solar panels and tests done on the satellite assemblies. 
Lastly, we outlined a method to analyze the impact of deployment failure scenarios, comparing orbital simulation results with power budget constraints to assess the impact of failure on satellite functionality.\\

\noindent The solar panel designs described in this paper will be utilized in the upcoming Ex-Alta~3 mission, scheduled for launch in 2025. 
To fulfill one of the main goals of this project, we will make the design files, assembly procedures, and best practices publicly available, facilitating the development of open-source satellites. 
Future work will focus on refining the design for enhanced performance, reducing risk, and improving the design of deployment mechanisms.
This work lowers the barrier of entry into space, makes satellite design easier and less expensive,  and promotes innovation and education in the space community - students helping students design better satellites.

\subsection*{Acknowledgements}
The authors graciously acknowledge the many members of AlbertaSat that contributed directly or indirectly to this work, including but not limited to D. G. Elliott, I. R. Mann, M. G. Lipsett, D. Sameoto, M. D. Fleishauer, K. S. Knudsen, O. Bauer, E. Saive, S. Wang, C. Lissinna, S. E. Damkjar, K. S. Ball, C. Nokes, C. Cupido, A. Martino, J. G. Grey, and T. Hrynyk. 
The authors would also like to express their gratitude towards L. Reeves and the CSDCMS for their unwavering dedication to creating opportunities for Canadian University students to pursue careers in aerospace. 
This work was undertaken with the financial support of the University of Alberta, the Natural Sciences and Engineering Research Council, and the Canadian Space Agency through the Canadian CubeSat Project. 
We thank M. T. Taschuk and G2V Optics Inc. for their kind support, allowing us to use a Sunbrick\texttrademark ~ Solar Simulator. 

%% file: tables/electricalCharTable.tex
\begin{table}[t]
\caption{Solar panel characteristics of Ex-Alta 2}
\label{tab:char}
\footnotesize \centering
\begin{tabular}{|c|c|c|c|c|}
\hline
\rowcolor[HTML]{C0C0C0} 
\multicolumn{1}{|c|}{\cellcolor[HTML]{C0C0C0}{\color[HTML]{000000} \textbf{category}}} & \cellcolor[HTML]{C0C0C0}{\color[HTML]{000000} \textbf{parameter}}                           & {\color[HTML]{000000} \textbf{min}} & {\color[HTML]{000000} \textbf{typical}} & {\color[HTML]{000000} \textbf{max}} \\ \hline
\rowcolor[HTML]{FFFFFF} 
\cellcolor[HTML]{C0C0C0}{\color[HTML]{000000} }                                        & {\color[HTML]{000000} MPP voltage (V)}                                                      & {\color[HTML]{000000} -}            & {\color[HTML]{000000} 14.4}             & {\color[HTML]{000000} -}            \\ \cline{2-5} 
\rowcolor[HTML]{FFFFFF} 
\cellcolor[HTML]{C0C0C0}{\color[HTML]{000000} }                                        & {\color[HTML]{000000} MPP current (A)}                                                      & {\color[HTML]{000000} -}            & {\color[HTML]{000000} 0.48}            & {\color[HTML]{000000} -}            \\ \cline{2-5} 
\rowcolor[HTML]{FFFFFF} 
\multirow{-3}{*}{\cellcolor[HTML]{C0C0C0}{\color[HTML]{000000} \textbf{output}}}       & {\color[HTML]{000000} orbital power (W)}                                                    & {\color[HTML]{000000} 5.2}          & {\color[HTML]{000000} 6.2}              & {\color[HTML]{000000} 7.4}          \\ \hline
\rowcolor[HTML]{FFFFFF} 
\cellcolor[HTML]{C0C0C0}{\color[HTML]{000000} }                                        & {\color[HTML]{000000} power cons. (mW)}                                                     & {\color[HTML]{000000} -}            & {\color[HTML]{000000} 60}               & {\color[HTML]{000000} 75}           \\ \cline{2-5} 
\rowcolor[HTML]{FFFFFF} 
\multirow{-2}{*}{\cellcolor[HTML]{C0C0C0}{\color[HTML]{000000} \textbf{operation}}}    & {\color[HTML]{000000} \begin{tabular}[c]{@{}c@{}}operating \\ temperature (C)\end{tabular}} & {\color[HTML]{000000} -40}          & {\color[HTML]{000000} -}                & {\color[HTML]{000000} 125} \\ \hline
\rowcolor[HTML]{FFFFFF} 
\cellcolor[HTML]{C0C0C0}{\color[HTML]{000000} }                                        & {\color[HTML]{000000} \begin{tabular}[c]{@{}c@{}}body-mounted \\ panel (g)\end{tabular}}    & {\color[HTML]{000000} 72}           & {\color[HTML]{000000} 82}               & {\color[HTML]{000000} 92}           \\ \cline{2-5} 
\rowcolor[HTML]{FFFFFF} 
\multirow{-2}{*}{\cellcolor[HTML]{C0C0C0}{\color[HTML]{000000} \textbf{mass}}}         & {\color[HTML]{000000} \begin{tabular}[c]{@{}c@{}}deployable \\ panel (g)\end{tabular}}      & {\color[HTML]{000000} 66}           & {\color[HTML]{000000} 76}               & {\color[HTML]{000000} 86}           \\ \hline 
\end{tabular}
\end{table}